# Some Recommended Protection Technologies for Cyber Crime Based on Social Engineering Techniques – Phishing

Wajeb GHARIBI,
Computer Science & Information Systems College, Jazan University,
Jazan, Kingdom of Saudi Arabia.
gharibi@jazanu.edu.sa

Abstract: *Phishing (password + fishing) is a form of cyber crime based on social engineering and site spoofing techniques. The name of 'phishing' is a conscious misspelling of the word 'fishing' and involves stealing confidential data from a user's computer and subsequently using the data to steal the user's money. In this paper, we study, discuss and propose the phishing attack stages and types, technologies for detection of phishing web pages, and conclude our paper with some important recommendations for preventing phishing for both consumer and company.*

## 1. INTRODUCTION

Nowadays, there is a huge variety of cyber threats that can be quite dangerous not only for big companies but also for ordinary user, who can be a potential victim for cybercriminals when using unsafe system for entering confidential data, such as login, password, credit card numbers, etc. Among popular computer threats it is possible to distinguish several types of them depending on the means and ways they are realized. They are: malicious software (malware), DDoS (Distributed Denial-of-Service) attacks, phishing, banking, exploiting vulnerabilities, botnets, threats for mobile phones, IP-communication threats, social networking threats and even spam. All of these threats try to violate one of the following criteria: confidentiality, integrity and accessibility. Lately, malicious software has turned into big business and cyber criminals became profitable organizations and able to perform any type of attack.

Phishing is a current social engineering attack that results in online identity theft. In a phishing attack, the attacker persuades the victim to reveal confidential information by using web site spoofing techniques. Social engineering is one of the strongest weapons in the armory of hackers and malicious code writers, as it is much easier to trick someone into giving his or her password for a system than to spend the effort to hack in. By 2007 social engineering techniques became the number-one method used by insiders to commit e-crimes, but unsuspecting users remain the predominant conduit for the authors of malicious code [1, 2].

It is unknown precisely how much phishing costs each year since impacted industries are reluctant to release figures; estimates range from US$1 billion to 2.8 billion per year [3].

The rest of our paper is organized as follows: Section 2 demonstrates the phishing attack stages and types. Section 3. Technologies for detection of phishing web pages. In section 4, we give some affective recommendations strategies against phishing. Conclusions have been made in section 5.

## 2. PHISHING ATTACK STAGES AND TYPES

Generally, social engineering malware reproduces through a variety of channels, including e-mail, social software, websites, portable storage devices, and mobile devices. There are several different ways of trying to drive users to a fake web site:

- **Spam e-mail**, spoofed to look like correspondence from a legitimate financial institution.

- **Hostile profiling**, a targeted version of the above method: the cyber criminal exploits web sites that use e-mail addresses for user registration or password reminders and directs the phishing scam at specific users (asking them to confirm passwords, etc.).

- **Install a Trojan** that edits the hosts file, so that when the victim tries to browse to their bank's web site, they are re-directed to the fake site.

- **'Spear phishing'**, an attack on a specific organization in which the phisher simply asks for one employee's details and uses them to gain wider access to the rest of the network. For traditional phishing sites, removing either the hosting website or the domain (if only used for phishing) is sufficient to remove a phishing site.

Traditional type of phishing attack is shown on Figure 1.

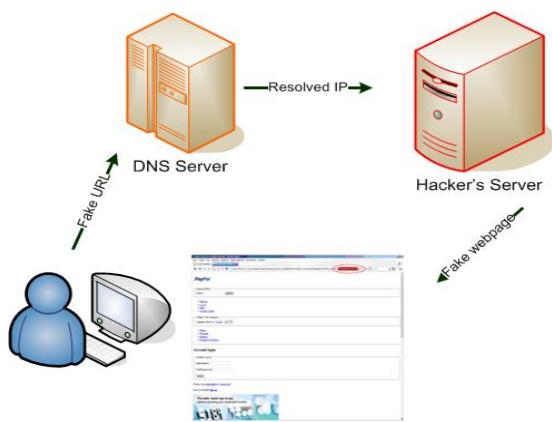

**Figure 1.** Traditional attack

Not all phishing attacks work in the manner just described. The 'rock-phish' gang3 has adapted its attack strategy to evade detection and maximize phishing-site availability. It has separated out the elements of the attack while adding redundancy in the face of take-down requests.

The gang first purchases a number of domain names with short, generally meaningless, names such as *lof80.info*. The email spam then contains a long URL such as *http://www.bank.com.id123.lof80.info/vr* where the first part of the URL is intended to make the site appear genuine and a mechanism such as `wildcard DNS' can be used to resolve all such variants to a particular IP address.

It then maps each of the domain names to a dynamic pool of compromised machines according to a gang-controlled name server. Each compromised machine runs a proxy system that relays requests to a back-end server system. This server is loaded with a large number (up to 20 at a time) of fake bank websites, all of which are available from any of the rock-phish machines. However, which bank site is reached depends solely upon the URL-path, after the first '/'. (Because the gang uses proxies, the real servers – that hold all the web pages and collate the stolen information – can be located almost anywhere.)

According to statistic presented by InfoSecurity more than 50% of all PhishTank reports are categorized as rock-phish (see Figure 2). Rock-phish domains and IPs also last longer than ordinary phishing sites: rockphish domains last for 95 hours on average while rock IPs last 172 hours, compared to 62 hours for regular phishing sites. These longer lifetimes occur despite impersonating around 20 banks simultaneously, which should draw the attention of more banks. One explanation for the longer lifetimes is that their attack method is not widely understood, leading to sluggish responses.

Splitting up the components of the phishing attack (domains, compromised machines and hosting servers) obfuscates the phishing behavior so that each individual decision maker (the domain registrar, ISP system administrator) cannot recognize the nature of the attack as easily when an impersonated domain name is used (such as barclaysbankk.com), or HTML for a bank site is found in a hidden sub-directory on a hijacked machine.

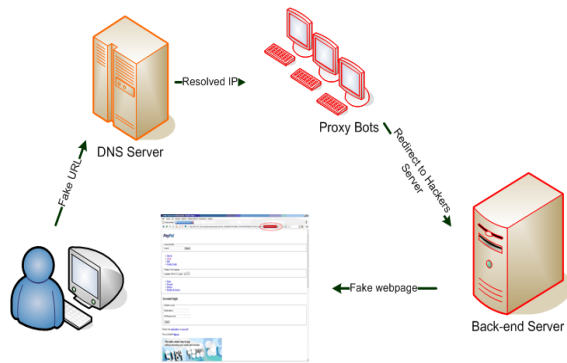

**Figure 2.** Rock-phish attack

Further innovation by the gang dubbed 'fast-flux' by the anti-phishing community (see Figure 3). It arranged for its domains to resolve to a set of five IP addresses for a short period, then switched to another five. This of course 'eats up' many hundreds of IP addresses a week (4572 addresses during our eight-week collection period), but the agility makes it almost entirely impractical to 'take down' the hosting machines. The gang is likely to have large numbers of compromised machines available (probably in the form of botnets), since if they are not used to serve up phishing websites, they are available for sending email spam. Fast-flux IP addresses remained alive for 139 hours on average, slightly less time than for rock-phish IPs. This is likely a reflection of the nature of the compromised hosts – consumer machines with dynamic IP address assignment – since the sites were not actively taken down. Domains were very long-lived (252 hours on average). This is because many fast-flux sites were not actually phishing sites at all. Instead, many were hosting mule-recruitment sites or selling diet pills and Viagra.

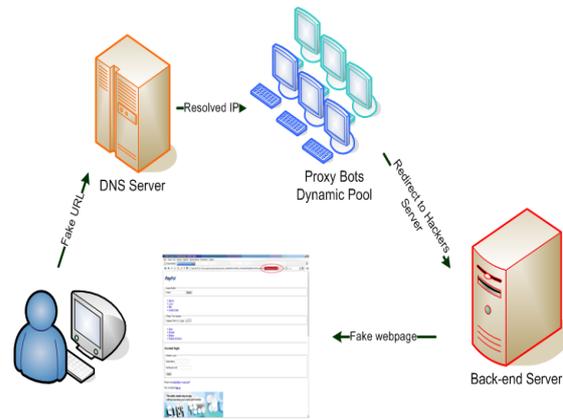

**Figure 3.** Fast-flux attacks

## 3. TECHNOLOGIES FOR DETECTION OF PHISHING WEBPAGES

Many anti-phishing solutions have been proposed to date. Some approaches attempt to solve the phishing problem at the e-mail level. Obviously, such techniques are closely related to anti-spam research. In fact, anti-spam techniques have proven to be quite effective in also intercepting phishing e-mails. Unfortunately, the effectiveness of anti-spam techniques often depends on many critical factors such as regular filter training and the availability of anti-spam tools and are currently not used by the majority of Internet users.

Well-known solutions in literature are SpoofGuard [4-6] and PwdHash [7-10]. SpoofGuard looks for phishing symptoms (e.g., obfuscated URLS) in web pages and raises alerts. Pwd-Hash, in contrast, creates domain-specific passwords that are rendered useless if they are submitted to another domain (e.g., a password for www.gmail.com will be different if submitted to www.attacker.com).

AntiPhish tool [11] takes a different approach and keeps track of where sensitive in-formation is being submitted. That is, if it detects that confidential information such as a password is being entered into a form on an untrusted web site, a warning is generated and the pending operation is canceled.

An interesting solution that has been proposed by Dhamija et al. [5] involves the use of a so-called dynamic security skin on the user's browser. The technique allows a remote server to prove its identity in a way that is easy for humans to verify, but difficult for phishers to spoof. The disadvantage of this approach is that it requires effort by the user. That is, the user needs to be aware of the phishing threat and check for signs that the site he is visiting is spoofed.

The most popular and widely-deployed techniques, however, are based on the use of blacklists of phishing domains that the browser refuses to visit [12-15].

## 4. RECOMMENDATIONS

In this section, we give some recommendations for phishing protection:

### 4.1 For a Company:

- Create corporate policies for E-mail content so that legitimate E-mail cannot be confused with phishing.

- Provide a right way and stronger authentication at web sites for the consumer to validate that the received E-mail is legitimate.

- Monitor the Internet for potential phishing web sites and implement good quality of anti-virus to filter and block known phishing sites at the gateway.

### 4.2 For a Consumer:

- Automatically block malicious E-mail by implementing Spam detectors which can help to keep the consumer from ever opening the suspicious E-mail.

- Automatically detect and delete malicious software and spyware by installing any of specialized commercial programs.

- Moreover, we recommend a combination of countermeasures that will minimize the number of phishing attacks delivered to consumers; increase the likelihood that the consumer will recognize a phishing attack; and minimize the opportunities for the consumer to inadvertently release sensitive information.

- Finally, education remains critical so consumers are aware of both the phishing techniques and how legitimate entities will communicate with them via E-mail and the web.

## 5. CONCLUSIONS

Con artists have been around for centuries, but E-mail and the World Wide Web provide them with the tools to reach thousands or millions of potential victims in minutes at almost no expense. With phishing attacks, con artists must still gain the consumer's confidence to be successful. So, the final technical solution to phishing involves significant infrastructure changes in the Internet that are beyond the ability of any one institution to deploy. However, there are many steps, as we mentioned, that can be taken to reduce the consumer's vulnerability to phishing attacks.